Directional Distribution of Ocean Surface Roughness Observed in Microwave Radar Backscattering

Paul A. Hwang

Remote Sensing Division, Naval Research Laboratory, Washington DC, USA

This work was supported by the Office of Naval Research (NRL PE 61153N). NRL publication number: NRL/JA/7260-15-0073.

P. A. Hwang is with the Naval Research Laboratory, 4555 Overlook Ave SW, Washington, DC 20375 USA (e-mail: paul.hwang@nrl.navy.mil)

*Abstract* — The directional distribution of ocean surface roughness is examined using the Ku, C and L band microwave radar backscattering. The parameters characterizing the upwind-downwind and upwind-crosswind variations show nonmonotonic dependence on wind speed based on the analysis of Ku, C and L band geophysical model functions (GMFs). A similarity relationship is derived from the GMFs of the three frequency bands to serve as the foundation of modeling the ocean surface roughness directional distribution function. The quantitative impacts on the magnitude and directional properties of the calculated radar backscattering cross section from using different directional distribution functions are investigated. The result indicates that it is important to include both upwind-downwind and upwind-crosswind variations in the directional distribution function in order to correctly model the radar scattering from the ocean surface.

# I. INTRODUCTION

Scattering and emission of microwaves from the ocean surface are greatly influenced by the ocean surface roughness properties. The ocean surface roughness is contributed mainly by short scale surface waves. In terms of the mean square slopes (MSS), more than 77% is from waves between 0.02 and 6 m wavelengths (referred to as the intermediate and short scale waves or ISW in this paper) for wind speeds less than 20 m/s [1]. The analysis is based on field measurements of ISW spectra using high-frequency response wave gauges installed on a free-drifting floating platform. The purpose of free-drifting operation is to alleviate the effect of Doppler frequency shift on deriving the wave number resolution from the frequency spectra computed from the surface wave time series [2, 3]. Subsequent investigations extend the coverage of long scale waves by incorporating our understanding of the wind sea equilibrium spectrum and short scale waves by integrating microwave measurements. The results again reconfirm the dominant role of ISW contributing to the ocean surface MSS [4].

With the common interest in surface roughness for remote sensing and air-sea interaction, the two scientific communities contribute considerable efforts to developing directional spectral models over the full length scales [1, 4-20]. Extensive reviews on the theoretical or empirical basis of the proposed spectral models have been given in the cited papers and they will not be repeated here. In essence, this is a demanding task because the important wavelength scales cover about six orders of magnitude ($10^{-3}$ to $10^{3}$ m), which corresponds to about 13 orders of magnitude in spectral magnitude [5]. No single measurement device is capable of providing such kind of broad-range coverage. Most models divide the wavenumber coverage into several segments. The spectral function in each wave number segment is formulated and then the multiple segments are stitched together by continuity matching of neighboring segments.

Short scale wave data are difficult to acquire in the field environment, so frequently laboratory results are employed in the model development. One exception to this rule is the H spectrum model [1, 4, 17, 18, 20], which depends on field data only, recognizing the conflicting similarity requirements in modeling

ISW in the laboratory environment. In particular, to simulate the gravity waves, the Froude number similarity needs to be observed. On the other hand, to simulate wind forcing, the Reynolds number similarity dictates. Extending to capillary and capillary-gravity length scales, the Weber number enters into the mix [21]. Translating the laboratory result into the equivalent field conditions thus becomes very difficult and usually not addressed when laboratory and field data are combined. Another unique feature of the H spectrum is that the formulation is anchored in the ISW band (0.02 to 6 m wavelengths); as discussed at the beginning of this section, the ISW band is the dominant contributor of the total mean square slope.

Following the discussion of the source function balance [22], the main effort of the H spectral model development is in searching of a similarity relationship of the dimensionless spectrum $B(k)$ as a function of the dimensionless wind forcing parameter $u_*/c$; where $k$ is the wave number, $c$ is the phase speed and $u_*$ is the wind friction velocity. A power-law similarity function

$$B(k) = A(k)(u_*/c)^{a(k)} \tag{1}$$

is eventually established with the wave spectra collected over several years of field measurements [2]. The similarity function becomes the foundation of the H spectrum model [1]. Asymptotic functions are formulated to extend the wavenumber coverage: toward the longer scale by assuming the equilibrium wind wave spectrum function, and toward shorter scales by relying on the empirical wind speed dependence of radar backscattering data [17, 18]. Later on, the Ku, C, and L band GMFs developed from global airborne and spaceborne scatterometer and SAR measurements are incorporated to further refine the H spectral model and to extend the wind speed coverage to hurricane conditions [4, 20]. It is gratifying that the results of the inversed spectral functions from the three GMFs are generally in good agreement with those derived from in situ measurements in low and moderate wind speeds ($U_{10}$ less than about 14 m/s). Because field data of ISW spectrum in high winds are not available, indirect verification of the roughness spectrum performance in high winds is based on comparing the microwave scattering and emission computations with field observations. For example, the *VV* (vertical transmit vertical receive) normalized radar cross section (NRCS) is dominated by the Bragg scattering of surface roughness. Using the H roughness spectrum, the calculated *VV* NRCS is found to be generally within -2 and +3 dB across Ku, C and L frequency bands for wind speeds up to 60 m/s and incidence angles between 20° and 50° using either the composite-surface Bragg scattering (CB) or the second order small slope approximation (SSA2) solution [20].

Here, the investigation of ocean surface roughness properties is continued with special focus on the directional distribution. Field measurements of the directional information of short waves are scarce and designs of directional distribution models frequently rely on extrapolating the results of long scale waves or laboratory data. Earlier attempts in the 1970s to 1980s to obtain the directional distribution of short surface wave components using radar backscattering results [5, 10] are hindered by the relatively meager quantity of available data during those early days of ocean remote sensing. The quantity and quality of remote sensing measurements have improved significantly since those attempts and it is overdue to revisit this important ocean remote sensing topic using the most updated radar backscattering results.

Anticipating that $u_*/c$ remains an important factor influencing the directional distribution of short scale wind generated waves, section II explores the similarity relationship of the directional distribution function $D(\phi) = f(u_*/c)$ using the Ku, C and L band GMFs to serve as the proxy of ocean surface roughness measurements. The directional model based on the similarity function, coupled with the 1D roughness spectra, is then used to compute the NRCS. Section III presents the result and discussion of NRCS calculations using different directional distribution functions and the comparison with Ku, C and L GMFs. Section IV is a summary.

## II. SIMILARITY RELATIONSHIP OF THE DIRECTIONAL COEFFICIENTS

*a. Method*

The directional distributions of GMFs are expressed as the sum of cosine harmonics under the assumption of symmetry with respect to the wind direction: $D(\phi) = \sum_0^N (X_n \cos n\phi)^x$; the number of harmonic terms $N$ varies from 2 (Ku2001 and CMOD5 family) to 5 (Ku2011 and L band Aquarius), the exponent $x$ is usually 1 but the CMOD5 family employs $x=1.6$ [23-26]. In this work, we convert those different VV GMFs to the common expression

$$\sigma_0(\phi) = B_0 D(\phi), \qquad (2)$$

$$D(\phi) = (1 + B_1 \cos\phi + B_2 \cos 2\phi). \qquad (3)$$

$B_0$ is thus the directionally integrated NRCS; the directional distribution includes the first two harmonic terms: $B_1$ characterizes the upwind-downwind asymmetry and $B_2$ characterizes the upwind-crosswind asymmetry and the directional distribution function $D(\phi)$ is normalized, that is, $\int_{-\pi}^{\pi} D(\phi) d\phi / 2\pi = 1$. The coefficients $B_0$, $B_1$ and $B_2$ can be readily obtained from the upwind, crosswind and downwind values of the NRCS: $\sigma_0(0)$, $\sigma_0(\pi/2)$, and $\sigma_0(\pi)$:

$$B_0 = \frac{\sigma_0(0) + \sigma_0(\pi) + 2\sigma_0(\pi/2)}{4}. \qquad (4)$$

$$B_1 = \frac{\sigma_0(0) - \sigma_0(\pi)}{2B_0}. \qquad (5)$$

$$B_2 = 1 - \frac{\sigma_0(\pi/2)}{B_0}. \qquad (6)$$

Our main task is to search for the similarity relationship of $B_1$ and $B_2$ making use of the Ku, C and L band GMFs.

At this point, it is appropriate to compare the directional distribution parameters observed by microwave scatterometers and those specified in different spectral models. Figure 1 shows an example of the $B_2$ parameter for several wave numbers corresponding to C (Figure 1a) and Ku bands (Figure 1b) at $\theta$=30, 40, 50 and 60°. The $B_2$ obtained from GMFs are shown with markers, illustrating an increasing trend in low to moderate winds then decreasing for higher winds. In each GMF, the peak value of $B_2$ increases for increasing wave number $k$ and the location of the peak shifts toward a lower wind speed as $k$ increases. The continuous curves are for the Elfouhaily (E) spectral model [15] in Figures 1a and 1b, and for the Donelan-Banner-Plant (DBP) spectral model [8, 11, 12] in Figures 1c and 1d. The E model prescribes a monotonically increasing $B_2$ (at least for $U_{10}$ greater than ~ 2 m/s) whereas the DBP model gives a monotonically decreasing result. Both models expect little variations in high wave number components, the threshold of diminished variation of $B_2(k)$ is about 250 rad/m for the E model, and about 400 rad/m for the DBP model. Clearly, there is plenty of room for improvement in the directional distribution function of roughness spectral models.

*b. Similarity relationship*

Numerical calculation shows that the *VV* backscattering is well represented by the simple Bragg scattering solution (without including the tilting effects) in the incidence angle ($\theta$) range between about 45° and 75°, e.g., see Figure 1 in [4]. For the purpose of establishing the similarity relationship, we examine the Ku, C and L band *VV* GMFs in this $\theta$ range. Table 1 lists the recommended ranges of wind speeds and incidence angles of the Ku, C and L band GMFs examined in this paper. Figure 2 shows a typical example of the $B_1$ and $B_2$ coefficients processed with the listed GMFs at $\theta$=45° and 53°. In this section, we make the assumption that the $B_1$ and $B_2$ of the GMFs are identical to those of the corresponding surface roughness components; in places where distinction between the two is necessary, as will be further elaborated in section III, the $B_1$ and $B_2$ are reserved for the GMFs, and the corresponding roughness quantities are given as $b_1$ and $b_2$.

As mentioned in section II.a, $B_1$ characterizes the upwind-downwind asymmetry and $B_2$ characterizes the upwind-crosswind asymmetry. Both factors can be considered as indices of surface wave nonlinearity. Because wave nonlinearity increases with wind speed, at $U_{10}=0$, $B_1=B_2=0$ is expected, which seems to be observed in the L band data but not in C and Ku bands, especially for the $B_1$ data. Historically, the $B_1$ measurements are more scattered and the results more difficult to interpret, e.g., see discussions in [27, 28]. For the L band results, we display both the measured data points (cyan squares) and the fitting (blue) curves reported in [29], showing the relatively sparse measurements and the diverse quality of data fitting. The non-decreasing trend of Ku and C band $B_1$ data and the non-vanishing $B_2$ in the low wind condition may be attributed to the relatively weak signals of surface returns (low wind speed) in high level of noise coming from non-local wind sources, such as background swell and ambient currents. These non-local sources can modify the properties of locally wind-generated short scale waves.

Despite the data scatter, taking into consideration the analysis given in the last paragraph there appears to be a general trend of non-monotonic wind speed dependence in both $B_1$ and $B_2$, with the peak of the dependence shifting toward higher wind speed as the Bragg wavenumber $k_B$ decreases. For reference, the Bragg wavenumbers for Ku and C bands at 45° and 53°, and L band at 45° are shown in the right panels of Figure 2. Interestingly, when the data are presented as a function of $u_*/c$, the peaks seem to converge toward $u_*/c$ near 1.5, and the maximum of each curve shows a generally decreasing trend as $k_B$ decreases. The drag coefficient used in this paper is equation 2 of [4]: $C_{10} = 10^{-5}\left(-0.16U_{10}^2 + 9.67U_{10} + 80.58\right)$, with a small modification such that $u_*$ increases monotonically with $U_{10}$: for $U_{10} > U_t$, $C_{10}$ is replaced by $C_{10t}\left(U_{10}U_t^{-1}\right)^\alpha$, where $C_{10t}$ is $C_{10}$ at $U_t$; we use $U_t$=45 m s$^{-1}$, $\alpha$=-1.5 thus in high wind speeds $u_*$ increases slowly with, and proportional to the square root of, $U_{10}$.

The top panels of Figure 3 show the wavenumber dependence of $B_{1max}$ and $B_{2max}$, least squares polynomial fittings yield:

$$B_{1\max} = -4.66 \times 10^{-7} k^2 + 4.55 \times 10^{-4} k + 6.78 \times 10^{-2}$$
$$B_{2\max} = -1.90 \times 10^{-6} k^2 + 1.47 \times 10^{-3} k + 3.48 \times 10^{-1}. \quad (7)$$

The normalized variables $B_{in}=B_i/B_{imax}$, where $i=1$ and 2, are shown in the bottom panels of Figure 3. The data can be approximately by the thick magenta curves in the figure:

$$B_{in}(\xi) = \exp\left\{-4\left[\log\left(\frac{\xi}{1.5}\right)\right]^2\right\}, \; \xi = u_*/c, \; i=1,2. \quad (8)$$

Equations 7 – 8 represents the empirical similarity function for the directional parameters $B_1$ and $B_2$ of short surface waves. Using a similar approach described in [15], the long wave portion is "patched" to the short wave $B_1$ and $B_2$ for the full spectrum application (i.e., $B_{1full}=B_1+ B_{1long}$; $B_{2full}=B_2+ B_{2long}$). In this paper, $B_{1long}=0$ and $B_{2long}=\tanh[4/(U_{10}/c)^{2.5}]$, which corresponds to the expression given in [30]. From this point on, the subscripts "full" and "long" for $B_1$ and $B_2$ will be dropped unless clarification is needed. To ensure that the resulting NRCS computation does not become negative, the maximum values of $B_1$ and $B_2$ are further limited to be unity.

Figure 4 shows an example of the proposed $B_1(k)$ and $B_2(k)$ for wind speeds 10, 20, 30 and 40 m/s. For comparison, the corresponding $B_2(k)$ of the E spectrum [15] is also illustrated (red curves in Figure 4b). The directional distribution function of the E spectrum is symmetric in the upwind-downwind direction and contains only the $B_2$ term:

$$D_E(\phi) = (1 + B_{2E} \cos 2\phi)/2\pi,$$
$$B_{2E}(k) = \tanh\left[a_0 + a_p\left(\frac{c}{c_p}\right)^{2.5} + a_m\left(\frac{c_m}{c}\right)^{2.5}\right]. \quad (9)$$
$$a_0 = \frac{\ln 2}{4}, \; a_p = 4, \; a_m = 0.13\frac{u_*}{c}, \; c_m = \left(\frac{2g}{k_m}\right)^{0.5} = 0.23\text{m/s}$$

Considerable differences between the two sets of curves are apparent, consistent with the results shown in Figure 1. For example, in the short wave region $k>\sim 60$ rad/m, the E model predicts increasing $B_2$ with wind speed, whereas the similarity function derived from GMFs shows an opposite trend.

### III. APPLICATION TO NRCS COMPUTATION

Here we conduct numerical experiments to investigate the effect of using different directional distribution functions on the NRCS computation. To clarify the presentation of the results, the directional distribution of the NRCS is given as

$$D_\sigma(\phi) = (1 + B_1 \cos\phi + B_2 \cos 2\phi), \tag{10}$$

and the directional distribution of the surface roughness is given as

$$D_r(\phi) = (1 + b_1 \cos\phi + b_2 \cos 2\phi), \tag{11}$$

The similarity functions (7 – 8) are used for $b_1$ and $b_2$ in the roughness directional distribution function. The composite-surface Bragg (CB) scattering solution described in [31] is used to compute the *VV* NRCS using the most recent iteration of the H spectrum [20] coupled with the directional distribution function (7, 8, 11). The purpose is to test the sensitivity of various changes in the directional distribution of Bragg resonance wave components as well as the tilting slope components on the computed NRCS result. The goal is to achieve as close as possible the observed upwind-downwind and upwind-crosswind variations as well as the magnitude between computed and observed NRCS GMFs.

*a. Directional distribution*

The NRCS of Ku, C and L band microwave frequencies are computed for wind speeds between 3 and 60 m/s and $\theta$ between 20 and 60°. For clarity of presentation, results from each frequency are shown in separate figures: Figures 5, 6 and 7, respectively, with very similar format. For each frequency the GMF is used as the "calibration standard" and displayed with blue curves.

Serving as a reference, the results using the E spectrum and Gaussian distribution of the tilting slopes are shown with red curves and labeled 'EG' in the legend. For the H directional distribution discussed in section 2.2, three variations are given in the figure:

(1) Forcing $b_1=0$, thus the directional distribution is upwind-downwind symmetric (equivalent to the E model design), Gaussian distribution of the tilting slopes, shown with black curves and labeled 'Hb$_2$ G' in the legend;

(2) Keeping both $b_1$ and $b_2$ terms, Gaussian distribution of the tilting slopes, shown with green curves and labeled 'Hb$_1$b$_2$ G' in the legend;

(3) Keeping both $b_1$ and $b_2$ terms, Gram-Charlier distribution of the tilting slopes [33], shown with cyan curves and labeled 'Hb$_1$b$_2$ GC' in the legend.

The upwind-crosswind result $B_2$ is examined first (Figures 5b, 6b and 7b), followed by the upwind-downwind result $B_1$ (Figures 5a, 6a and 7a). Several key observations are summarized here:

(a) Comparing to the reference computation using the E directional distribution function (9) – red curves – the nonmonotonic trend of $B_2(U_{10})$ as observed in the GMF – blue curves – is improved considerably for all frequencies using the directional distribution (11) established on the empirical similarity functions (7, 8), even when $b_1$ is forced to be 0 such that the roughness distribution is upwind-downwind symmetric as specified in the E roughness model.

(b) Comparing the two sets of curves computed with $D_r(\phi)$ (11) and the Gaussian distribution of the tilting slopes, the one considering both $b_1$ and $b_2$ terms, shown with green curves, yields much better agreement with the GMF (blue curves) than the one considering the $b_2$ term alone (black curves) for Ku and L bands (Figures 5b and 7b) but the result is mixed for the C band, depending on the wind speed range (Figure 6b). The sharp increase of the L band GMF for $U_{10}$ greater than about 25 m/s, shown with continuous curves in Figure 7b, is likely the artifact of high order polynomial function given in [36]; such sharp increase is not found in the GMF design of [29], shown as blue circles and available only for $\theta=45°$.

(c) Comparing the two sets of curves computed with $D_r(\phi)$ (11) with both $b_1$ and $b_2$ term, the results differ only slightly between using the Gaussian distribution (green curves) and the Gram-Charlier distribution (cyan curves) of the tilting slopes. The database for the Gram-Charlier PDF is limited to $U_{10}$ less than 14 m/s [33] and extrapolation of the Gram-Charlier PDF beyond that wind speed is not warranted.

(d) With Gaussian PDF for the tilting slopes, all variations of the directional distribution tested here show very small $B_1$ in the NRCS output. The magnitude of $B_1$ is raised considerably with the Gram-Charlier PDF (cyan curves), but the computed results are significant different from that of the GMFs (blue curves) in all frequencies (Figures 5a, 6a and 7a).

*b. NRCS magnitude*

The comparison of the calculated NRCS magnitudes with Ku, C and L band GMFs has been discussed extensively [4, 20]. Particularly, in [20] the computed VV NRCS using the second order small slope approximation (SSA2) and composite surface Bragg (CB) models are compared with the Ku, C, and L band GMFs. For a wind range of wind speeds ($U_{10} \leq 60$ m/s for Ku and C bands and 35 m/s for L band) and incidence angles ($20 \leq \theta \leq 50°$), the H roughness spectrum produces agreement with GMFs to within mostly ±2dB for Ku and L bands and -2 to +3dB for C band. Here the focus is on the impact of the directional distribution of the surface roughness spectral model, and also looking into improving the C band computation.

Figure 8a presents the comparison result for the L band Aquarius data obtained at 3 incidence angles (29.4, 38.4 and 46.3°), shown with blue circles, pluses and triangles, respectively; the maximum wind speed is about 35 m/s [32]. The computed NRCS are shown with continuous curves. For clarify, only three different sets of computations are illustrated: (a) directional distribution with only the $b_2$ term and

Gaussian PDF for tilting slopes (labeled $b_2$G); (b) directional distribution with both $b_1$ and $b_2$ terms, and Gaussian PDF for tilting slopes (labeled $b_1b_2$G); and (c) directional distribution with both $b_1$ and $b_2$ terms, and Gram-Charlier PDF for tilting slopes (labeled $b_1b_2$GC). Considerable differences are found between computations using the $b_2$ term only vs. those with both $b_1$ and $b_2$ terms. In comparison, the difference between specifying Gaussian PDF vs. Gram-Charlier PDF for tilting slopes is relatively small.

Figure 8b shows the C band results for $\theta$=30, 40, 50 and 60°. The CMOD5.n results are illustrated with blue circles, pluses, triangles and squares, the CMOD5.h results are illustrated with cyan color with identical markers; the difference between the two GMFs may reach slightly less than 1 dB at high winds. The computed NRCS are shown with continuous curves for three different variations, all with Gaussian PDF for tilting slopes: (a) directional distribution with only the $b_2$ term, high-wind switch [4] turned off (labeled h0$b_2$G); (b) directional distribution with both $b_1$ and $b_2$ terms, high-wind switch turned off (labeled h0$b_1b_2$G); and (c) directional distribution with both $b_1$ and $b_2$ terms, high-wind switch turned on (labeled h2$b_1b_2$G). For C band computation, turning on the high wind switch yields poor results in high winds and high incidence angles ($U_{10}\geq$~15 m/s, $\theta\geq$~50°). With the high-wind switch off, the computed NRCS using the directional distribution with only $b_2$ term produces slightly better results than that with both $b_1$ and $b_2$ terms.

The high wind switch is used to accommodate the observed change in the $u_*/c$ exponent in the similarity relationship (1) for wind speeds ranging from mild to hurricane conditions, as observed in the radar spectrometer analysis of short surface waves described in Figure 3a of reference [4]. The exponent is given a fixed number 0.75 for $u_*/c\geq 3$ ($U_{10}\geq 16$ m/s for C band) in [4], which is derived from the NRCS computations without accounting for the relative permittivity modification by breaking entrained air. When the air modification of relative permittivity is considered, the surface reflectivity decreases and the exponent needs to be increased to 1.0, as can be evaluated from combining Figure 3a of reference [4] and

Figure 6 of reference [20]. The two-branch design of the $u_*/c$ exponent is in fact a coarse approximation of the observed evolution of the wind speed exponent in high wind conditions. As shown in Figure 3a of reference [4], the variation of the $u_*/c$ exponent in high winds is continuous and wave number dependent. This complicated behavior is not currently fully assimilated in the H spectral model.

Figure 8c shows the Ku band results for $\theta$=30, 40, 50 and 60°. The Ku2001 GMF is illustrated with blue circles, pluses, triangles and squares, the Ku2011 GMF (for $\theta$=53° only) is illustrated with cyan triangles. In high winds, Ku2011 at $\theta$=53° is higher than Ku2001 at $\theta$=50°, illustrating the still evolving and uncertain nature of all the GMFs, especially in high winds. The computed NRCS are shown with continuous curves for three different variations, all with Gaussian PDF for tilting slopes: (a) directional distribution with only $b_2$ term, high-wind switch turned off (labeled h0b$_2$G); (b) directional distribution with only $b_2$ term, high-wind switch turned on (labeled h2b$_2$G); and (c) directional distribution with both $b_1$ and $b_2$ terms, high-wind switch turned on (labeled h2b$_1$b$_2$G). For Ku band computation, turning on the high wind switch yields considerably better results in high winds and all incidence angles ($U_{10} \geq \sim 15$ m/s). With the high-wind switch on, the computed NRCS using directional distribution with only $b_2$ term produces slightly better results than that with both $b_1$ and $b_2$ terms in high winds but slightly worse in low winds.

Figure 9 shows the ratio between GMFs and computed NRCS $\delta\sigma_{0VV}$, results shown with black curves in Figure 8, visually judged to be in best agreement with the GMFs, are used for this presentation (Figures 9a, 9b and 9c for Ku, C and L bands, respectively). The difference between computed NRCS and GMFs is mostly within ±2 dB for $U_{10}$ up to 60 m/s and $\theta$ between 20 and 60° (except for L band, the $\theta$ of the reported Aquarius data is between 29.4 and 46.3°). Figures 9d, 9e and 9f are partially reproduced from Figure 18 in [20], in which the NRCS computation uses the 1D H roughness spectral model coupled with the E directional distribution function. There is a small improvement in the results using the directional distribution described in this paper for all frequencies. For the C band, the bigger difference is caused by the high-wind switch, which is turned on in the computation reported in [20] (Figure 9e) but turned off in Figure 9b (also see the discussion above for Figure 8b).

## IV. Discussion and Summary

The earliest documented attempt to obtain the directional distribution of short surface wave components using radar backscattering results seems to be by Pierson and Stacy [5]. The effort was continued by Donelan and Pierson [10]. Despite the valiant efforts, the two publications are 127 and 59 pages respectively, the available data was simply too limited during those early days of microwave remote sensing of the ocean.

After 40 plus years since the first effort and with many more airborne and satellite missions the data volume has expanded extraordinarily. Ocean vector wind retrieval using active and passive microwave measurements has become a routine task. Although the GMFs are still evolving, especially for high wind conditions, it is overdue for revisiting the directional properties of short surface waves using the expanded results of radar backscattering measurements.

In this work, a similarity relation is derived for the first two harmonics of the directional distribution function of short surface waves using the Ku, C and L band GMFs (section II); the first harmonic $b_1$ characterizes the upwind-downwind variation and the second harmonic $b_2$ characterizes the upwind-crosswind variation. The results derived from GMFs are quite different from those specifies in published spectral models (e.g., Figures 1 and 4b).

The result from numerical experiments using different variations of the directional distribution function shows that including the $b_1$ term (upwind-downwind variation) in the roughness spectrum model produces better agreement of the upwind-crosswind variation with the radar cross section observations as summarized in the GMFs (Figures 5b, 6b and 7b). On the other hand, based on our current understanding the backscattering is contributed from Bragg resonance of both advancing and receding wave components, therefore, even though the ocean surface roughness has upwind-downwind asymmetry, if the tilting surface slopes are symmetrically distributed (such as the Gaussian distribution) the radar backscattering is expected to be upwind-downwind symmetric. This is confirmed with the numerical experiments (Figures 5a, 6a and 7a). Specifying a non-Gaussian PDF for tilting slopes (e.g., Gram-Charlier) is able to generate upwind-downwind asymmetry in the computed NRCS but the results are very different from those observed in the GMFs (Figures 5a, 6a and 7a). It seems that the source of upwind-downwind asymmetry of radar backscattering remains somewhat a mystery.

The magnitude of the computed NRCS is also influenced by the directional distribution function (Figure 8) but the effect is relatively small compared to other uncertainties of the roughness spectral models. One example shown in section III.b is the ($u_*/c$) exponent of the similarity relationship (1). Over all, the NRCS computed with the directional distribution function (11) established with the similarity relationship (7, 8) produces significant improvement in the backscattering directional properties, especially $B_2$ (Figures 5 – 7), and incremental improvement in the NRCS magnitude (Figure 9).

## VI. LIST OF TABLES

Table 1. Stated ranges of wind speed and incidence angle of the GMFs examined in this paper

| VV GMF | $U_{10}$ range (m/s) | $\theta$ range (°) |
|---|---|---|
| Ku2011 | 0~70 | 53 |
| Ku2001 | 0~70 | 16~66 |
| CMOD5.h | 0.5~50 | 18~58 |
| CMOD5.n | 0.5~50 | 18~58 |
| L Aquarius | 0~35 | 29.4, 38.4, 46.3 |
| L *Yueh et al.* [2010] | 4~28 | 45 |

## VII. LIST OF FIGURES

Figure 1. The $B_2$ parameter observed in (a) C and (b) Ku band GMFs at $\theta$=30, 40, 50 and 60°, shown with circles, pluses, triangles and squares, respectively; the corresponding $B_2$ values of the E spectral model [15] are illustrated with continuous curves (solid, dashed, dashed-dotted, and dotted); (c) and (d) show the $B_2$ variation of the DBP spectral model [8, 11, 12] at selected wave numbers specified in the legends.

Figure 2. The first two harmonics of the directional distribution function: (a, c) $B_1$, and (b, d) $B_2$, derived from Ku, C and L band GMFs for the incidence angles indicated in the legends of the left panels. The corresponding Bragg resonance wave numbers are shown in the legends of the right panels. For the L band data of [29] (Y10), the reported fitting curves and the original data used for the fitting are both displayed to illustrate the quality of $B_1$ and $B_2$ observations and curve fitting. In general, $B_1$ is a more difficult measurement, see text for further discussion.

Figure 3. Derivation of the similarity relationship of $B_1$ and $B_2$ given in normalized form by the maximum value (a) $B_{1\max}(k)$ and (b) $B_{2\max}(k)$, respectively; the normalized coefficients (c) $B_{1n}=B_1/B_{1\max}$, and (d) $B_{2n}=B_2/B_{2\max}(k)$ are approximated by $B_{in}(\xi) = \exp\{-4[\log(\xi/1.5)]^2\}$, $\xi=u_*/c$ and $i$=1 and 2.

Figure 4. (a) $B_1$ and (b) $B_2$ as a function of $k$ and $U_{10}$ obtained from the similarity relation given in Figure 3. In (b) the corresponding result of the E spectrum model [15] is also displayed.

Figure 5. Comparison of (a) $B_1$ and (b) $B_2$ between Ku2001 *VV* GMF and CB model computation using the 1D H roughness spectrum coupled with different directional distribution functions and tilting slope PDFs, see text for additional explanation; $\theta$=30, 40, 50 and 60°.

Figure 6. Same as Figure 5, except for comparison with C band CMOD5.n.

Figure 7. Same as Figure 5, except for comparison with L band Aquarius data at $\theta$=29.4, 38.4 and 46.3°.

Figure 8. Comparison of the NRCS magnitude of (a) L, (b) C, and (c) Ku band GMFs and CB solutions using the 1D H roughness spectrum coupled with different directional distribution functions and tilting

slope PDFs, see text for additional explanation.

Figure 9. The ratio between GMFs and computed NRCS $\delta\sigma_{0VV}$, results shown with black curves in Figure 8, visually judged to be in better agreement with GMFs, are used for this presentation: (a) Ku, (b) C and (c) L bands; (d), (e) and (f) are the corresponding results based on the 1D H roughness spectrum coupled the directional distribution function of [15] and presented in [20].

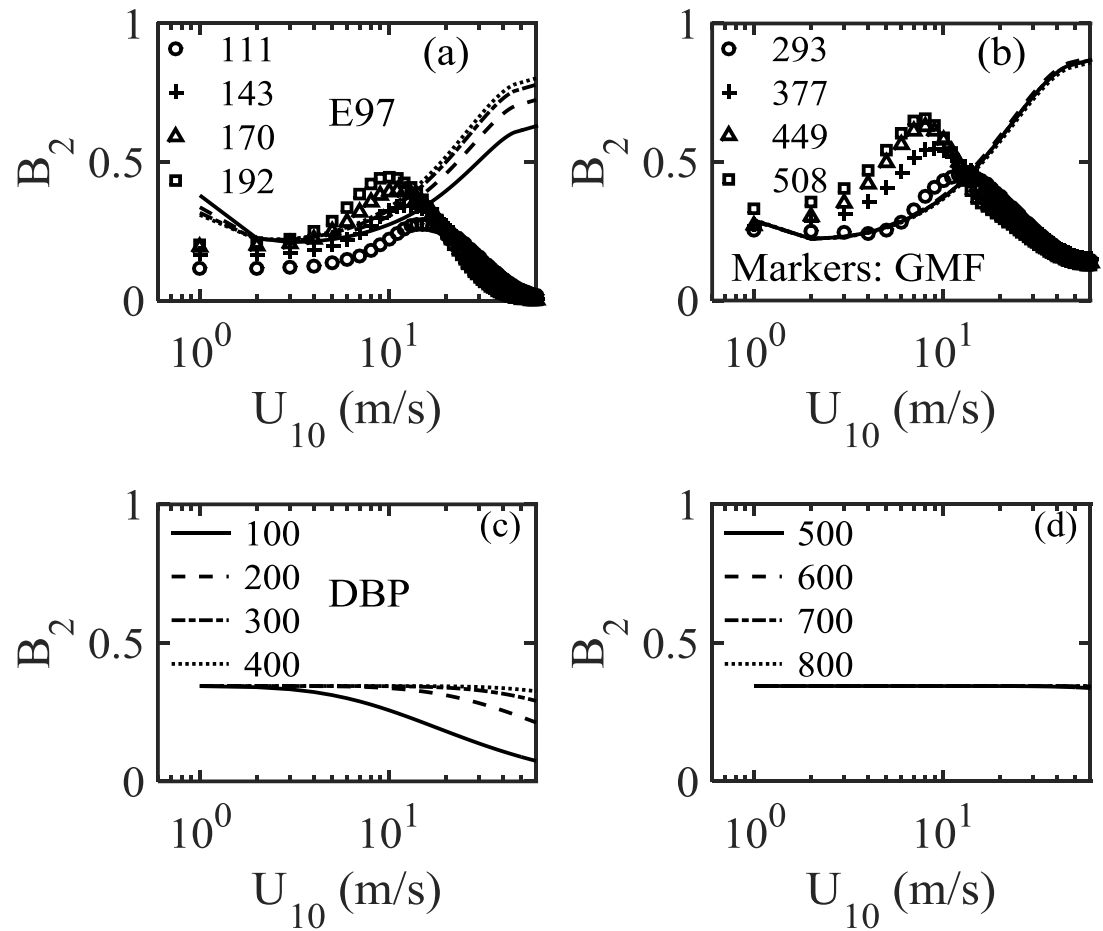

Figure 1. The $B_2$ parameter observed in (a) C and (b) Ku band GMFs at $\theta$=30, 40, 50 and 60°, shown with circles, pluses, triangles and squares, respectively; the corresponding $B_2$ values of the E spectral model [15] are illustrated with continuous curves (solid, dashed, dashed-dotted, and dotted); (c) and (d) show the $B_2$ variation of the DBP spectral model [8, 11, 12] at selected wave numbers specified in the legends.

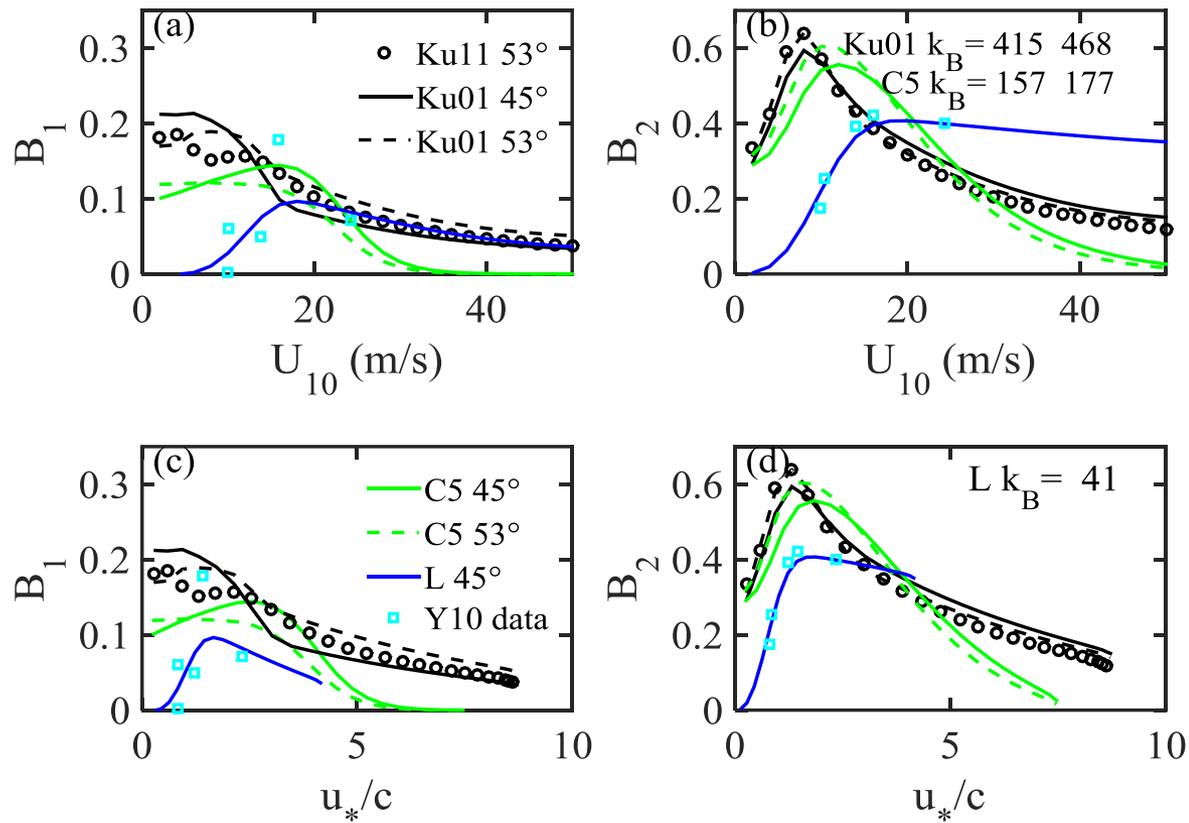

Figure 2. The first two harmonics of the directional distribution function: (a, c) $B_1$, and (b, d) $B_2$, derived from Ku, C and L band GMFs for the incidence angles indicated in the legends of the left panels. The corresponding Bragg resonance wave numbers are shown in the legends of the right panels. For the L band data of [29] (Y10), the reported fitting curves and the original data used for the fitting are both displayed to illustrate the quality of $B_1$ and $B_2$ observations and curve fitting. In general, $B_1$ is a more difficult measurement, see text for further discussion.

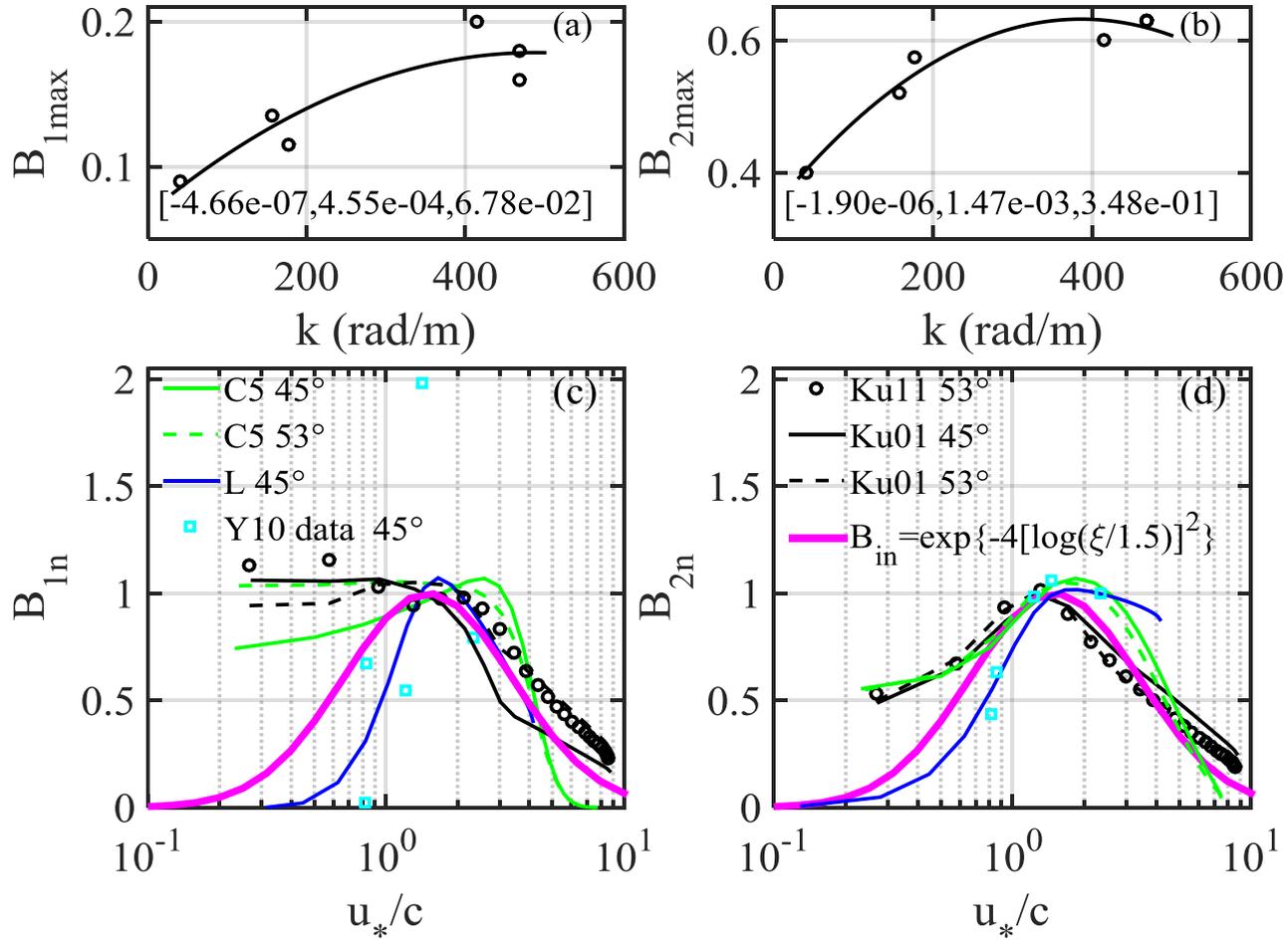

Figure 3. Derivation of the similarity relationship of $B_1$ and $B_2$ given in normalized form by the maximum value (a) $B_{1max}(k)$ and (b) $B_{2max}(k)$, respectively; the normalized coefficients (c) $B_{1n}=B_1/B_{1max}$, and (d) $B_{2n}=B_2/B_{2max}(k)$ are approximated by $B_{in}(\xi)=\exp\{-4[\log(\xi/1.5)]^2\}$, $\xi=u_*/c$ and $i=1$ and 2.

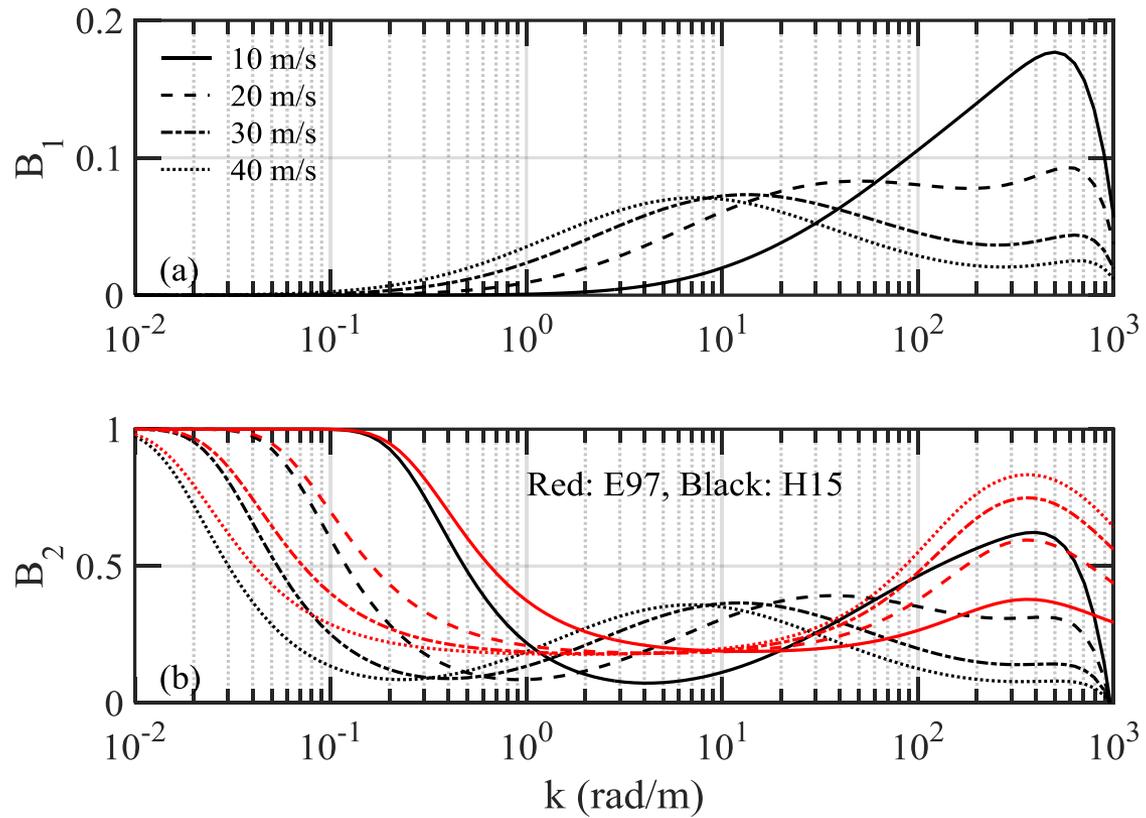

Figure 4. (a) $B_1$ and (b) $B_2$ as a function of $k$ and $U_{10}$ obtained from the similarity relation given in Figure 3. In (b) the corresponding result of the E spectrum model [15] is also displayed.

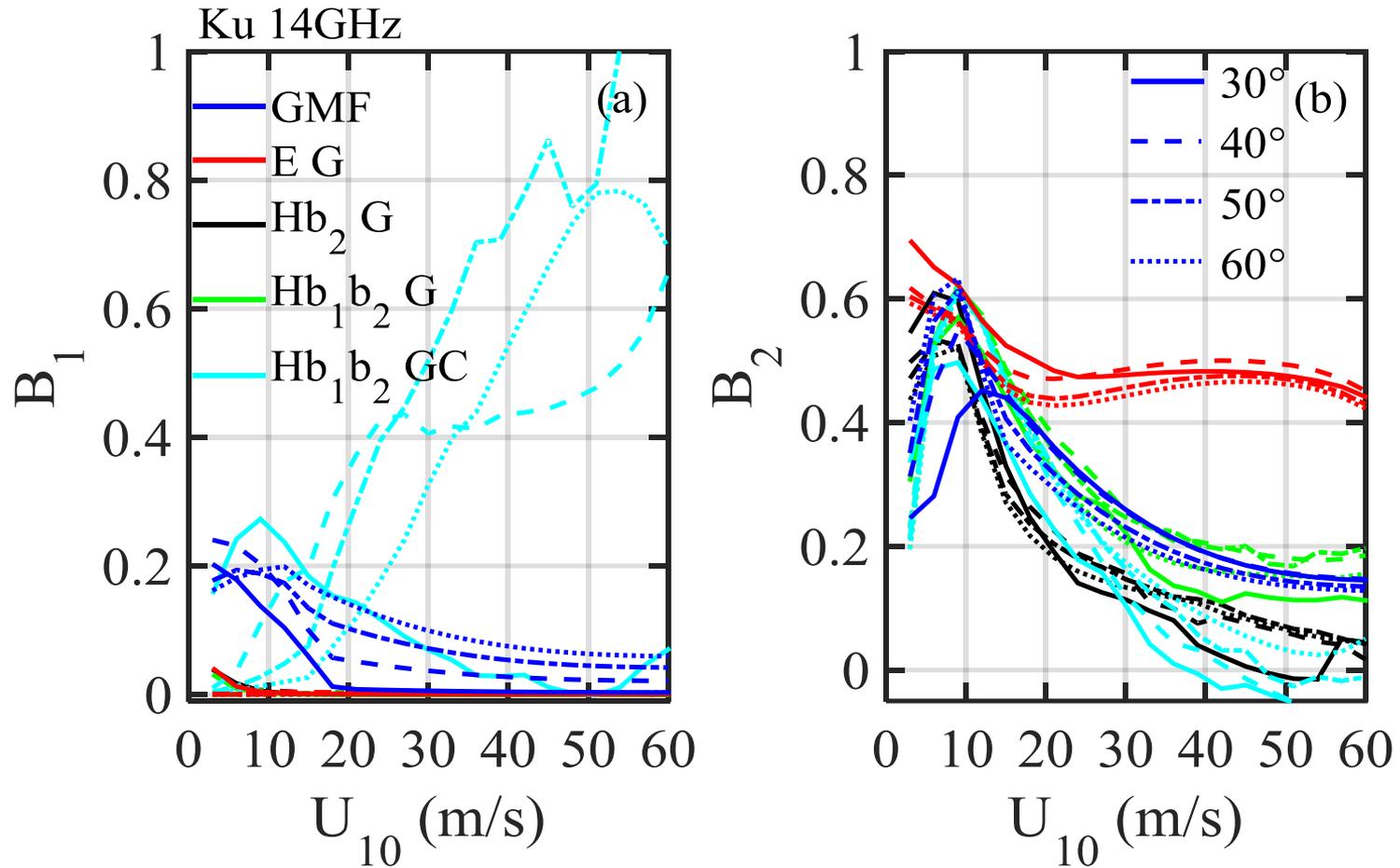

Figure 5. Comparison of (a) $B_1$ and (b) $B_2$ between Ku2001 *VV* GMF and CB model computation using the 1D H roughness spectrum coupled with different directional distribution functions and tilting slope PDFs, see text for additional explanation; $\theta$=30, 40, 50 and 60°.

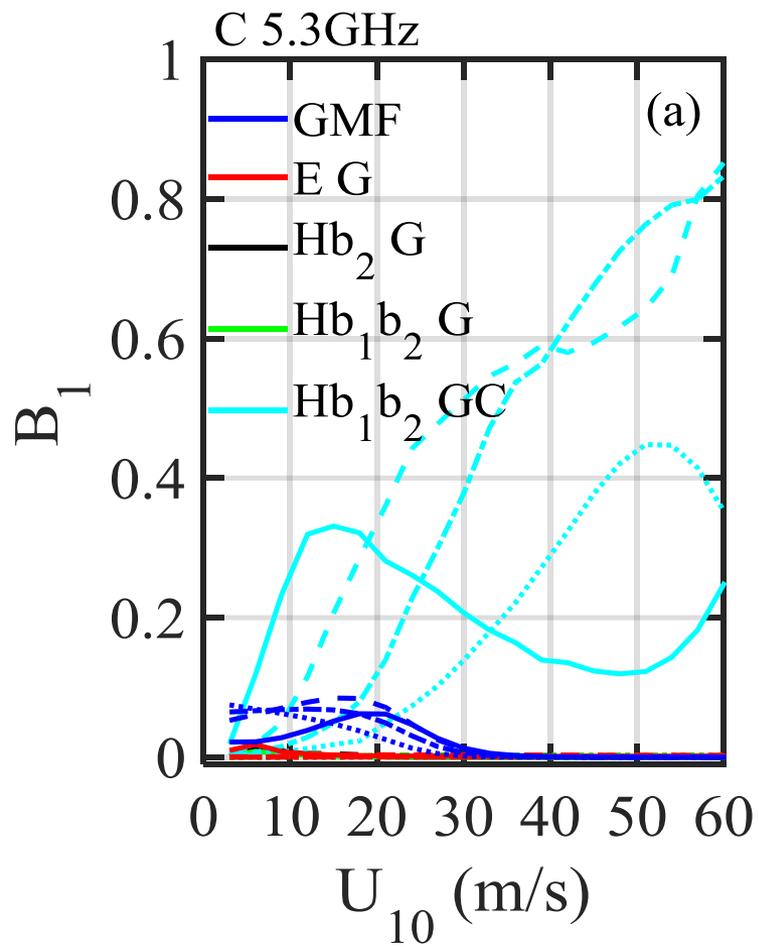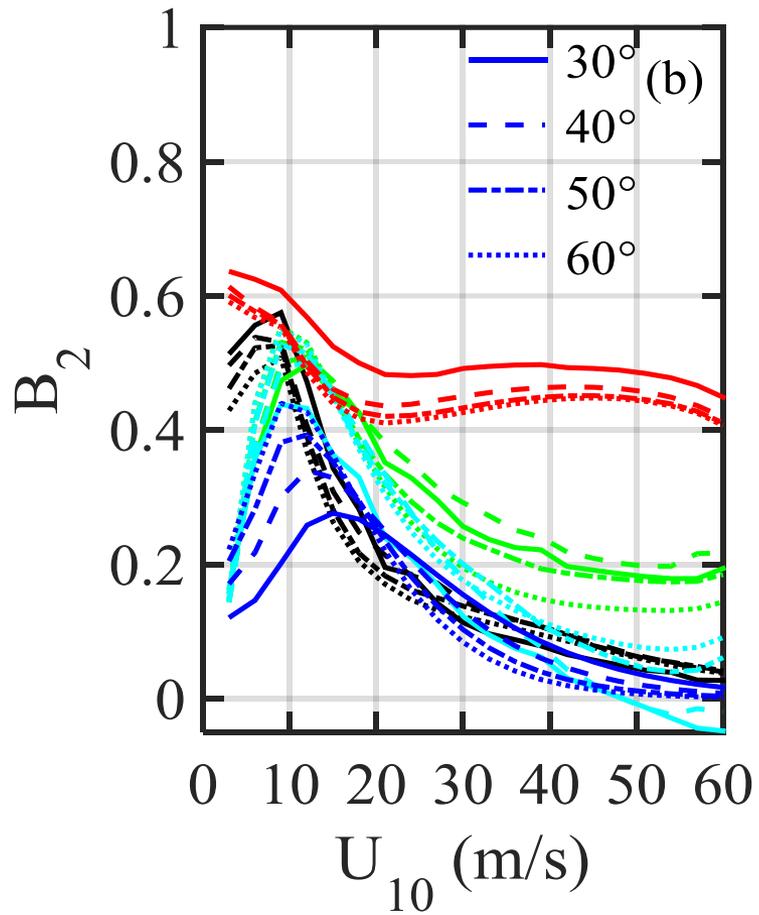

Figure 6. Same as Figure 5, except for comparison with C band CMOD5.n.

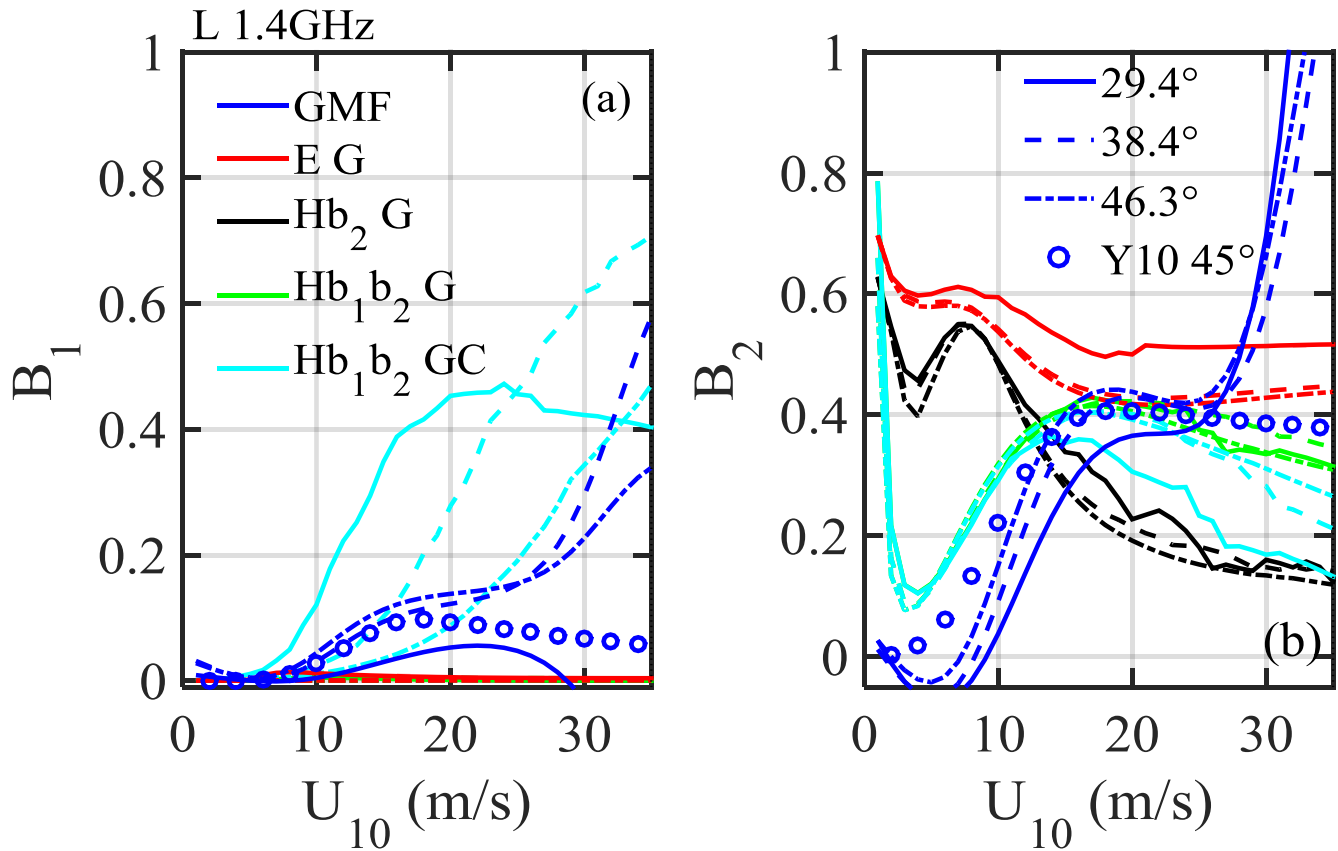

Figure 7. Same as Figure 5, except for comparison with L band Aquarius data at $\theta$=29.4, 38.4 and 46.3°.

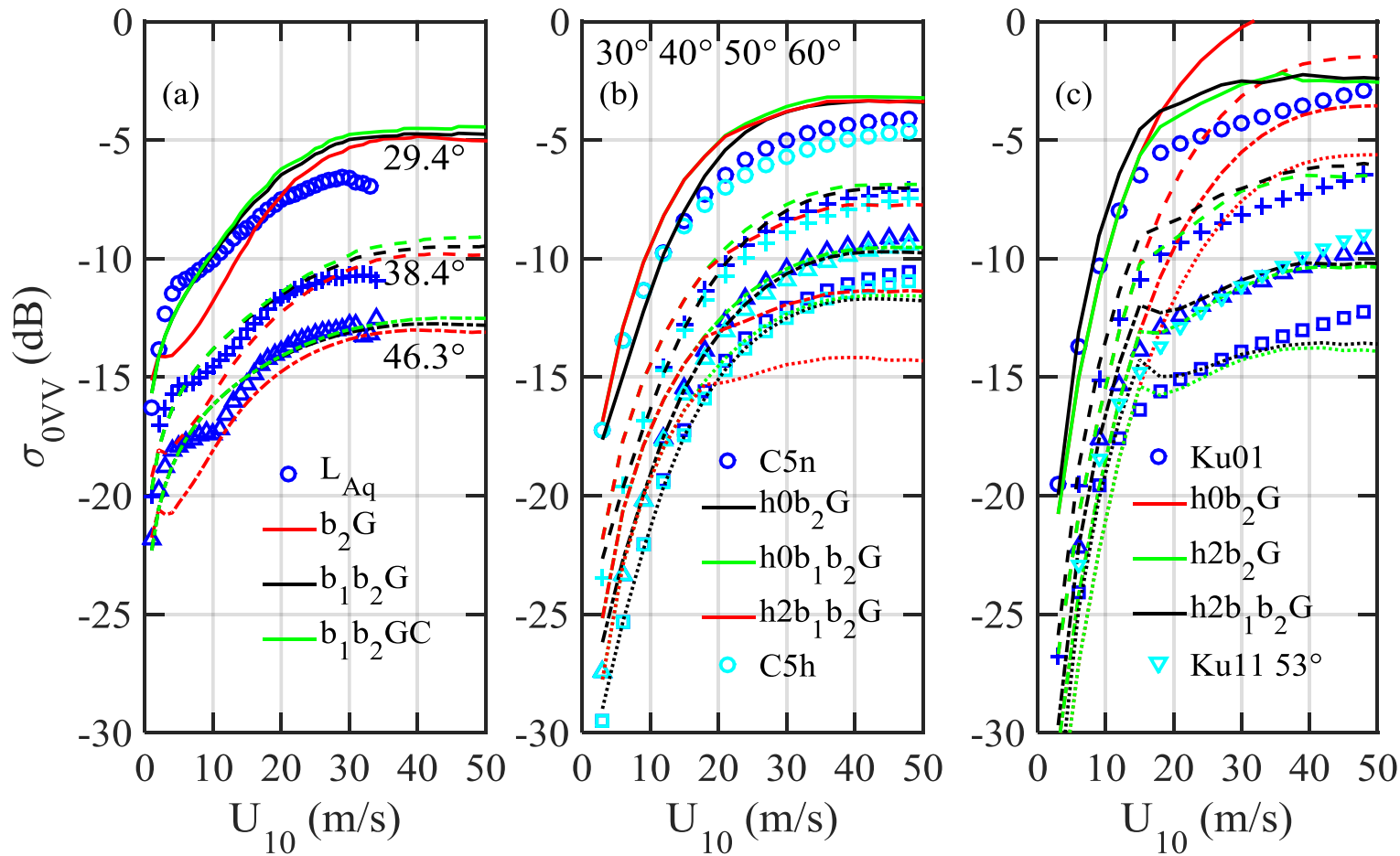

Figure 8. Comparison of the NRCS magnitude of (a) L, (b) C, and (c) Ku band GMFs and CB solutions using the 1D H roughness spectrum coupled with different directional distribution functions and tilting slope PDFs, see text for additional explanation.

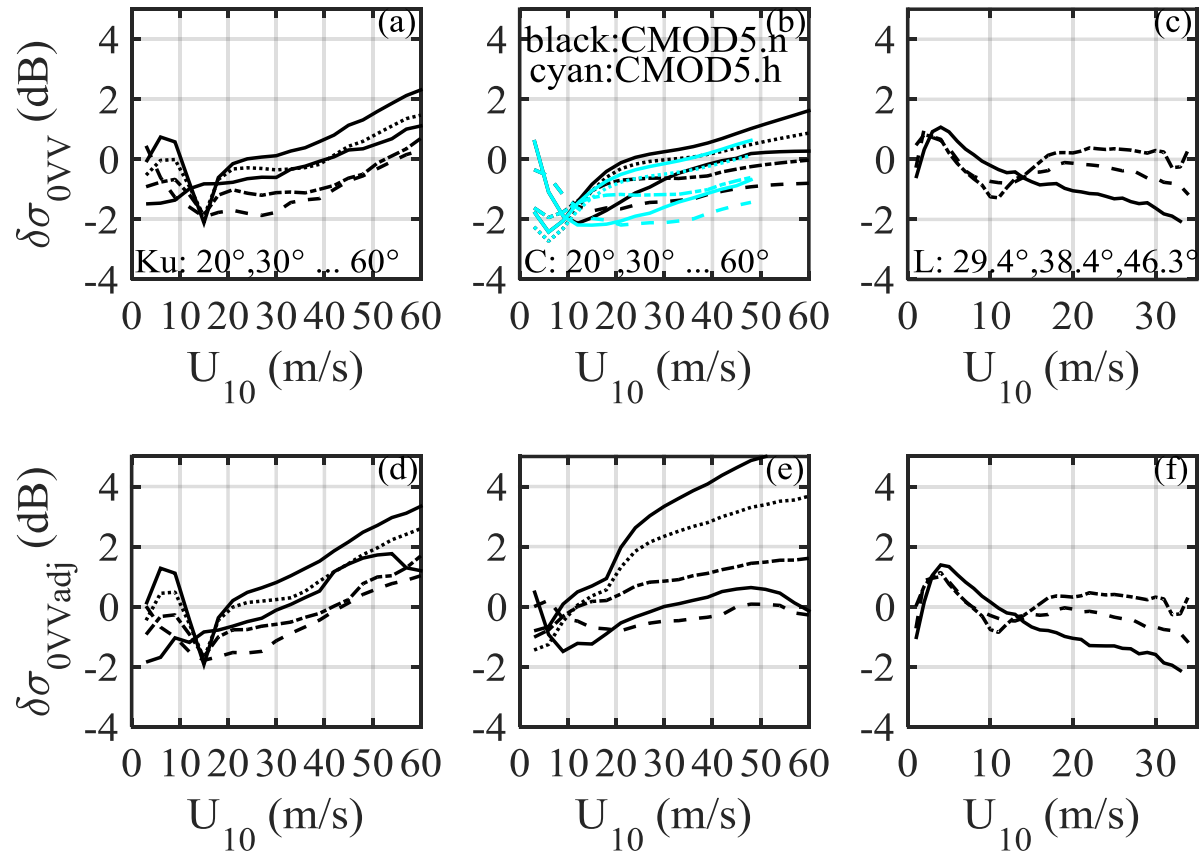

Figure 9. The ratio between GMFs and computed NRCS $\delta\sigma_{0VV}$, results shown with black curves in Figure 8, visually judged to be in better agreement with GMFs, are used for this presentation: (a) Ku, (b) C and (c) L bands; (d), (e) and (f) are the corresponding results based on the 1D H roughness spectrum coupled the directional distribution function of [15] and presented in [20].